\documentclass[aps,pra,twocolumn,superscriptaddress]{revtex4-2}
\usepackage{graphicx,epsf}
\usepackage{amsmath,amssymb}
\usepackage{tabularx}
\usepackage{color}

\newcommand{\dis}{U} 

\newcommand{\Cmax}{C_{\rm max}}
\newcommand{\Ccr}{C_{\rm crit}}

\graphicspath{{./}{./figs/}}

\begin{document}
	
\title{
Strain tuning of highly frustrated magnets:\\
Order and disorder in the distorted kagome Heisenberg antiferromagnet
}

\author{Mary Madelynn Nayga}
\affiliation{Institut f\"ur Theoretische Physik and W\"urzburg-Dresden Cluster of Excellence ct.qmat, Technische Universit\"at Dresden,
	01062 Dresden, Germany}
\affiliation{Max-Planck-Institut f\"ur Chemische Physik fester Stoffe, N\"othnitzer Str. 40,
	01187 Dresden, Germany}
\author{Matthias Vojta}
\affiliation{Institut f\"ur Theoretische Physik and W\"urzburg-Dresden Cluster of Excellence ct.qmat, Technische Universit\"at Dresden,
	01062 Dresden, Germany}

\date{\today}

\begin{abstract}
	Strain applied to a condensed-matter system can be used to engineer its excitation spectrum via artificial gauge fields, or it may tune the system through transitions between different phases. Here we demonstrate that strain tuning of the ground state of otherwise highly degenerate frustrated systems can induce novel phases, both ordered and disordered.
	For the classical Heisenberg antiferromagnet on the kagome lattice, we show that weak triaxial strain reduces the degeneracies of the system, leading to a classical spin liquid with  non-coplanar configurations, while stronger strain drives the system into a highly unconventional state which displays signatures of both spin-glass behavior and magnetic long-range order. We provide experimentally testable predictions for the magnetic structure factor, characterize the ground-state degeneracies and the excitation spectrum, and analyze the influence of sample shape and boundaries.
	Our work opens the way to strain engineering of highly frustrated magnets.
\end{abstract}
\maketitle


\section{Introduction}
\label{sec:intro}

The manipulation of many-body systems by external stimuli is widely used in both the search for novel phenomena and the realization of applications. Among the possible tools, pressure and the resulting lattice deformations are particularly appealing, as they do not introduce disorder -- as opposed to chemical substitution -- and they can either preserve or modify in a controlled fashion the lattice symmetries of the underlying system. Recent experimental progress in applying uniaxial  or otherwise inhomogeneous forces has led to the notion of straintronics, where specific strain patterns enable to engineer states and functionalities of novel materials \cite{pereira09,si_rev,naumis_rev}.

Examples of strain manipulation include the mechanical switching of nano-electronic graphene devices \cite{fogler08}, the occurrence of strain-induced Landau levels in graphene, with a spacing corresponding to ultra-large magnetic fields \cite{guinea09,levy10,voz_rev10}, the creation of artificial gauge fields for ultracold atoms and photonic crystals \cite{cold_gauge,aidelsburger18}, the proposals to realize Landau levels for emergent charge-neutral excitations in solids, such as magnons or Majorana spinons in quantum antiferromagnets \cite{rachel16a,nayga19}, and the modification of multi-component superconducting states \cite{hicks17,klauss21}.
In essentially all of these examples, one starts from a unique microscopic state which is modified by strain, changing either its static properties or its excitation spectrum.

In this paper, we extend the concept of strain tuning to highly degenerate many-body systems: Here, applying strain can be expected to have a singular effect, i.e., even small strain modifies the system's properties in a qualitative fashion.
We choose to discuss the effect of inhomogeneous strain applied to highly frustrated antiferromagnets where degenerate ground-state manifolds result from competing interactions. We show that suitably chosen strain patterns can be used to induce particular forms of magnetic order as well as novel spin-liquid regimes, thus opening a new arena for strain-based engineering of states of matter.
Specifically we consider the classical Heisenberg antiferromagnet on the kagome lattice. Its highly degenerate ground state features Coulombic spin correlations, and it displays remarkably complex order-by-disorder phenomena at finite low temperature \cite{chalker92,ritchey93,harris92,huse92,zhito08,chern13}. Its spin-1/2 cousin is a prime candidate to realize a quantum spin-liquid ground state \cite{sachdev92,white11,balents12,messio12,iqbal13,sheng15,trebst16,normand17,pollmann17,wietek19}.
Here we focus on the effect of triaxial strain which partially preserves discrete lattice symmetries of the kagome lattice, Fig.~\ref{fig:setup}, while partially relieving strong geometric frustration. Increasing strain lifts the classical degeneracies of the unstrained system, first deforming the spin liquid into a non-coplanar one with pronounced spin correlations at $\vec Q=0$. Larger strain induces a transition into a highly unusual state, being connected with the inability to independently minimize the energy on every triangle. This state displays characteristics of a spin glass, but at the same time its magnetic structure factor shows sharp peaks corresponding to $\vec Q=0$ long-range order. While details of the ground-state configurations and low-energy excitation modes depend on the sample shape and boundaries, the gross features of the strained magnetic state appear robust. We connect our findings to known results for homogeneous uniaxial strain applied to the kagome Heisenberg model, and we comment on the role of quantum effects.

The remainder of the paper is organized as follows:
In Sec.~\ref{sec:model} we introduce the inhomogeneously strained Heisenberg model and discuss the re-writing of its Hamiltonian as sum of complete squares.
Sec.~\ref{sec:config and SF} describes the results for spin configurations and the spin structure factor, obtained from minimizing the classical energy.
Sec.~\ref{sec:E-landscape} then discusses the complex energy landscape at finite strain, implying glassy features, and the properties of the low-energy excitations.
A summary and discussion of open questions closes the paper.

As an aside, we note that the effect of strain on a kagome-lattice tight-binding model has been recently studied in Ref.~\onlinecite{liu20}, with focus on single-particle pseudomagnetic fields. Also, spontaneous (instead of imposed) distortions which relieve frustration in highly frustrated magnets have been discussed in earlier papers \cite{ueda00,tscherny02,smerald19}.


\section{Model and constraints}
\label{sec:model}

\subsection{\label{sec:kagome} Kagome Heisenberg model}

\begin{figure}[t!]
	\includegraphics[width=\columnwidth]{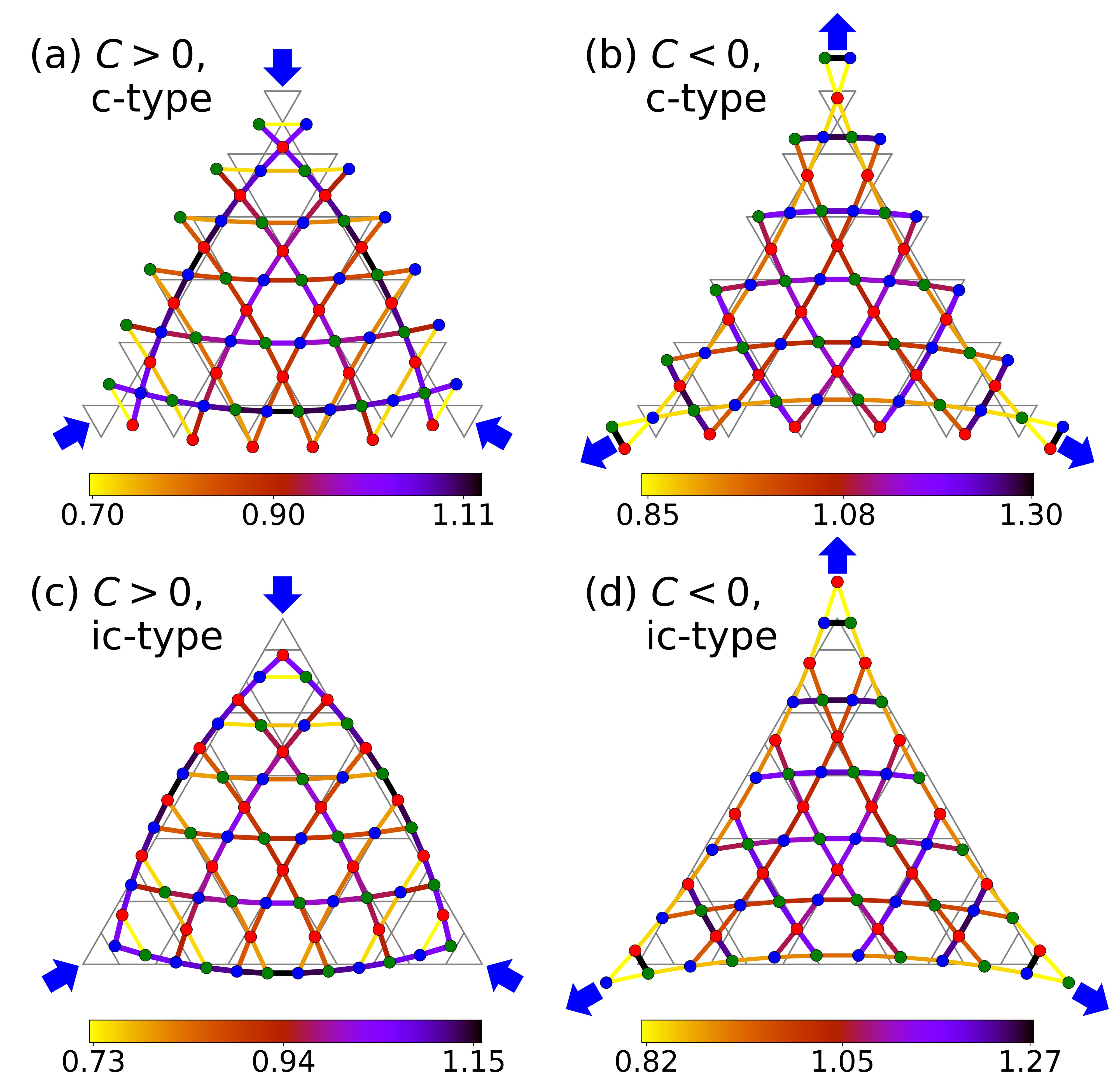}
	\includegraphics[width=0.95\columnwidth]{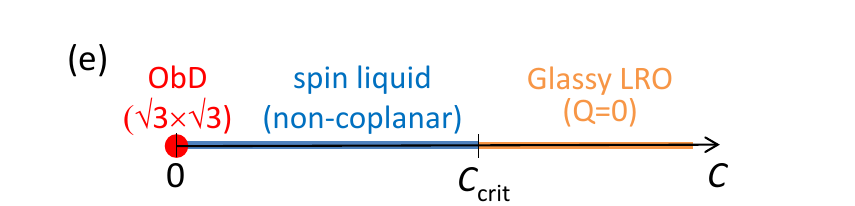}
	\caption{
		(a-d): Distorted kagome-lattice antiferromagnet, with displacements from triaxial strain, Eq.~\eqref{displace}, for $\beta=1$ and $|\bar{C}|=0.025/a_0$; the undistorted lattice is shown in light gray. Longer (shorter) bonds correspond to weaker (stronger) exchange couplings $J_{ij}$ as indicated by the color code. Arrows indicate the force applied to the sample.
		Panels (a,b) show the system with c-type boundaries, panels (c,d) with ic-type boundaries.
		Left (a,c) and right (b,d) panels correspond to positive and negative strain, respectively.
		The linear system size is $N=6$.
		(e) Schematic phase diagram of the distorted kagome-lattice antiferromagnet in the limit $T\to0$, for details see text.
	}
	\label{fig:setup}
\end{figure}
We consider a nearest-neighbor antiferromagnetic Heisenberg model on the kagome lattice, formed by corner-sharing triangles, with spatially varying couplings:
\begin{equation}
	\label{hk}
	\mathcal{H} = \sum_{\langle ij\rangle} J_{ij} \vec{S}_i \cdot \vec{S}_j\,.
\end{equation}
We exclusively focus on the classical case and treat the $\vec{S}_i$ as unit vectors. The homogeneous system, $J_{ij}\equiv J$, features a ground-state manifold which includes both coplanar and non-coplanar states
\cite{chalker92}. Finite-temperature fluctuations tend to select coplanar states via an order-by-disorder mechanism \cite{chalker92,ritchey93,harris92,huse92}, and the low-temperature regime displays weak long-range spin order corresponding to a $\sqrt{3}\times\sqrt{3}$ ordering pattern \cite{zhito08,chern13}.


\subsection{Triaxial strain}
\label{sec:triaxial}

Strain engineering goes back to the discussion of electron-phonon coupling in carbon nanotubes \cite{suzuura02} where strain-induced modulations of hopping matrix elements can emulate a vector potential for electrons \cite{guinea09}.
Central to our work is the modification of magnetic exchange couplings due to strain. In the distorted lattice, each magnetic ion is characterized by a displacement vector $\vec{\dis}_i$. This results in exchange couplings between neighboring ions, entering the Hamiltonian \eqref{hk},  which we assume to follow
\begin{equation}
	\label{ourjij}
	J_{ij} = J \left[ 1- \beta (|\vec{\delta}_{ij}|/a_0 - 1)\right]
\end{equation}
where $a_0$ is the reference bond length and  $\vec{\delta}_{ij}=\vec{R}_i+\vec{\dis}_i-\vec{R}_j-\vec{\dis}_j$ the length of the distorted bond. The materials parameter $\beta$ encodes the strength of magneto-elastic coupling. Realistic values of $\beta$ are in the range $1\ldots10$. For instance, the hopping matrix elements $t$ of graphene display a bond-length dependence with $\beta_t\approx 2\ldots 3$ \cite{naumis_rev}; for exchange couplings following $J=t^2/U$ where $U$ is an on-site Coulomb repulsion this would mean $\beta\approx 4\ldots 6$.
Note that Eq.~\eqref{ourjij} represents a linear approximation to the full (typically exponential) bond-length dependence of the exchange constant; assuming an exponential dependence yields qualitatively similar results as shown in the Appendix. Most numerical results are shown for $\beta=1$.

In the following, we focus on triaxial strain where the displacement vector is given by \cite{guinea09,peeters13}
\begin{equation}
	\label{displace}
	\vec{\dis}(x,y) = \bar{C} \big(2xy, x^2-y^2 \big)^T
\end{equation}
with $\bar{C}$ encoding the distortion amplitude, and we employ $\vec{\dis}_i=\vec{\dis}(\vec{R}_i)$. The dimensionless parameter
$C = \bar{C}\beta a_0$
specifies of the modulation strength of the $J_{ij}$.
The distortions described by Eq.~\eqref{displace} increase linearly with increasing distance from the sample center. Structural stability then requires to consider finite-sized samples.
Combining Eqs.~\eqref{ourjij} and \eqref{displace} we define -- for fixed sample size -- a maximum strain $\Cmax$ beyond which the outermost couplings become formally negative due to the linearization in Eq.~\eqref{ourjij}. For $\beta=1$ this means that the longest (i.e. weakest) bond takes twice its original length at maximum strain.
This maximum strain is inversely proportional to the linear system size, $\Cmax\propto 1/N$, therefore the thermodynamic limit $N\to\infty$ cannot be taken at fixed strain. As our results show, it is instead meaningful to consider the thermodynamic limit at fixed $C/\Cmax$, i.e., $N\to\infty$ with $(C N)$ fixed.
Moreover, we will sometimes refer to the combined limit $\beta\to\infty$ and $\bar C\to 0$, keeping $C = \bar{C}\beta a_0$ fixed, which reduces non-linearities in the strain dependence of couplings \cite{rachel16b}. In fact, the results at fixed $C$ depend only weakly on $\beta$ for $\beta>20$.

In order to partially preserve the discrete lattice symmetries, we primarily consider samples of triangular shape \cite{rachel16b}, Fig.~\ref{fig:setup}. Here, discrete rotation and mirror symmetries exist w.r.t. the sample center. For such samples, positive and negative strain, $\bar{C}\gtrless0$, correspond to qualitatively distinct distortion patterns, and we will display results for both.

Since finite-size properties will depend on the structure of the edges, we consider different types of edges: Kagome-lattice frustration is best preserved for edges with complete kagome triangles (dubbed c-type), Fig.~\ref{fig:setup}(a,b). As a representative for different edges, we choose those with all outer spins removed such that the outward triangles are incomplete (dubbed ic-type), Fig.~\ref{fig:setup}(c,d).
In both cases, we denote the linear sample size by $N$ where $N$ counts the number of complete triangles along a sample edge. Then, the total number of spins is $N_s = (3/2) N(N+1)$. Depending on $(N\mod 3)$ the center of the sample is either formed by an elementary triangle or hexagon.

For $C>0$ there are three weakest bonds located in the sample corners. For c-type edges we find $\Cmax^+ = \sqrt{3} / (4N-3)$ and for ic-type edges $\Cmax^+ = \sqrt{3} / (4N-9)$, both valid for arbitrary $\beta$.
For $C<0$ there are now six weakest bonds in the corner triangles. $\Cmax$ is given by a lengthy expression which is not particularly enlightening. However, in the limit $\beta \rightarrow \infty$ it simplifies to $|\Cmax^-| = \sqrt{3} / (2N-3)$ for both c-type and ic-type edges.


\subsection{Constraint satisfiability and critical strain}
\label{sec:critical}

The homogeneous Heisenberg Hamiltonian on the kagome lattice can be written as sum of complete squares, and this rewriting can be generalized to inhomogeneous couplings \cite{bilitewski17}: For each triangle $\alpha$ with spins $ijk$ we can define $\gamma_{i\alpha} = (J_{ij} J_{ik} / J_{jk})^{1/2}$ such that the Hamiltonian reads $\mathcal{H} = (1/2) \sum_\alpha \vec{L}_\alpha^2 + {\rm const}$, with $\vec{L}_\alpha = \sum_{i\in\alpha} \gamma_{i\alpha} \vec{S}_{i}$.

With small triaxial strain applied, the $\gamma_{i\alpha}$ will weakly deviate from their unstrained reference value unity, such that the minimization constraint $\vec{L}_\alpha=0$ can be fulfilled for all triangles of a finite sample. In contrast, for larger strain the $\gamma_{i\alpha}$ do no longer fulfill the triangle inequality for triangles $\alpha$ near the sample corners or edges, depending on the sign of $C$. This change defines a critical value of strain, $\Ccr^\pm$, where $\pm$ correspond to positive and negative strain, respectively.

While these considerations strictly apply to samples with c-type edges, samples with ic-type edges contain bonds not belonging to triangles, rendering the system less frustrated. Hence, the nature of the ground-state manifold depends on the type of sample edges. However, this difference turns out to be of minor importance for the magnetic structure factor for sufficiently large samples.

\begin{figure}[tb]
	\includegraphics[width=\linewidth]{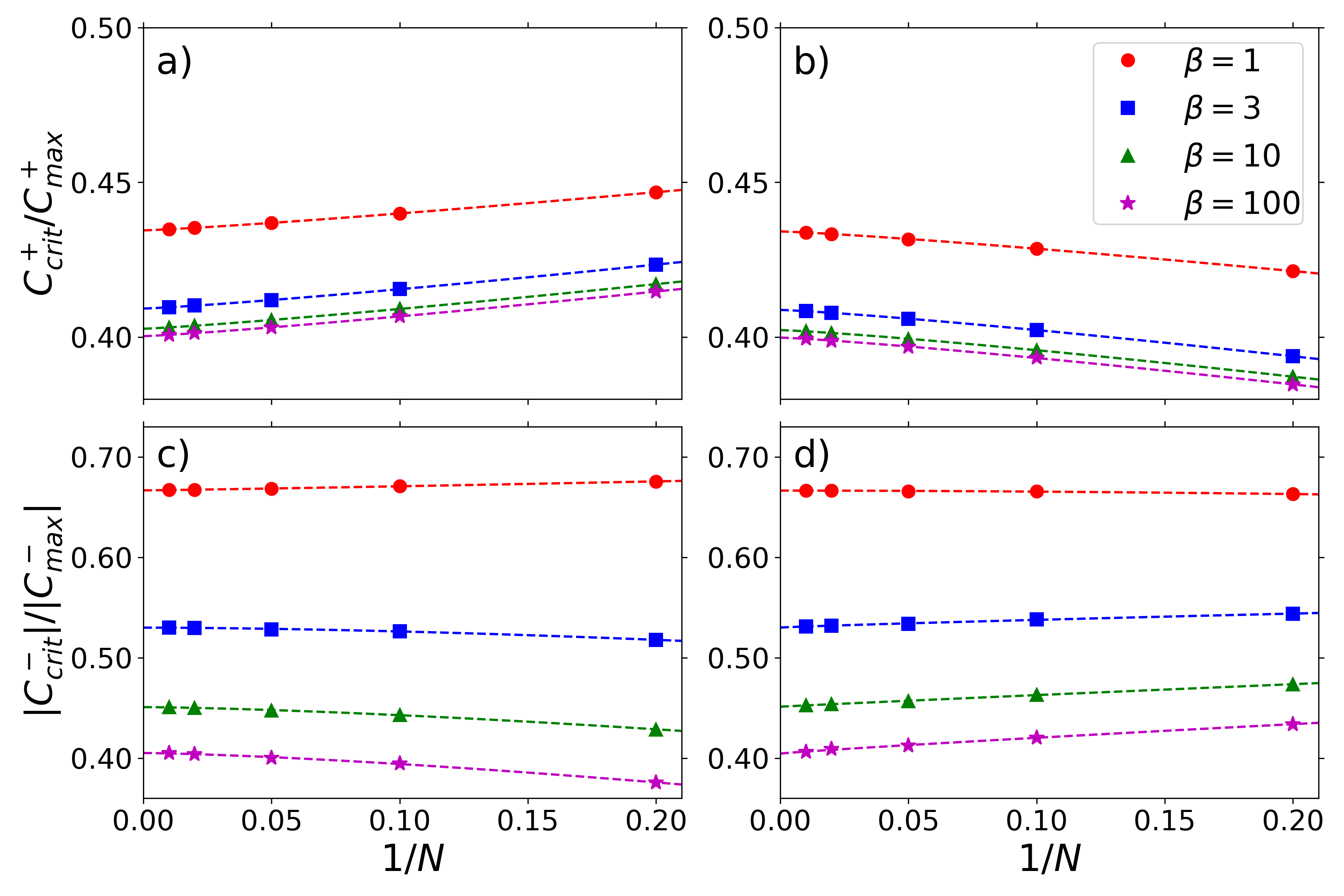}%
	\caption{
		Critical value of triaxial strain, plotted as $\Ccr^\pm/\Cmax^\pm$, as function of inverse linear system size, $1/N$, for different $\beta$ for (a,b) positive strain and (c,d) negative strain, both for samples with (a,c) c-type boundaries and (b,d) ic-type boundaries. $\Ccr/\Cmax$ approaches the value $0.4$ in the limit $\beta\to\infty$, $N\to\infty$ in all cases, for details see text.
	}
	\label{fig:ccr}
\end{figure}

Numerical results for the ratio of critical and maximum strain, $\Ccr/\Cmax$, for triangular samples are shown in Fig.~\ref{fig:ccr}. This quantity displays a mild dependence on system size $N$, but a stronger dependence on the magnetoelastic coupling $\beta$. For $|C|<|\Ccr^\pm|$ the strained system is strongly frustrated, with $\vec{L}_\alpha=0$ $\forall\alpha$ defining a degenerate manifold of liquid-like states \cite{bilitewski17,origami}, while for $|C|>|\Ccr^\pm|$ the constraint $\vec{L}_\alpha=0$ cannot be fulfilled for all triangles. Then, the condition $\sum_\alpha \vec{L}_\alpha^2\to\min$ induces tendencies to magnetic order, as we will see in the next section.

\begin{figure}[tb]
	\includegraphics[width=0.9\linewidth]{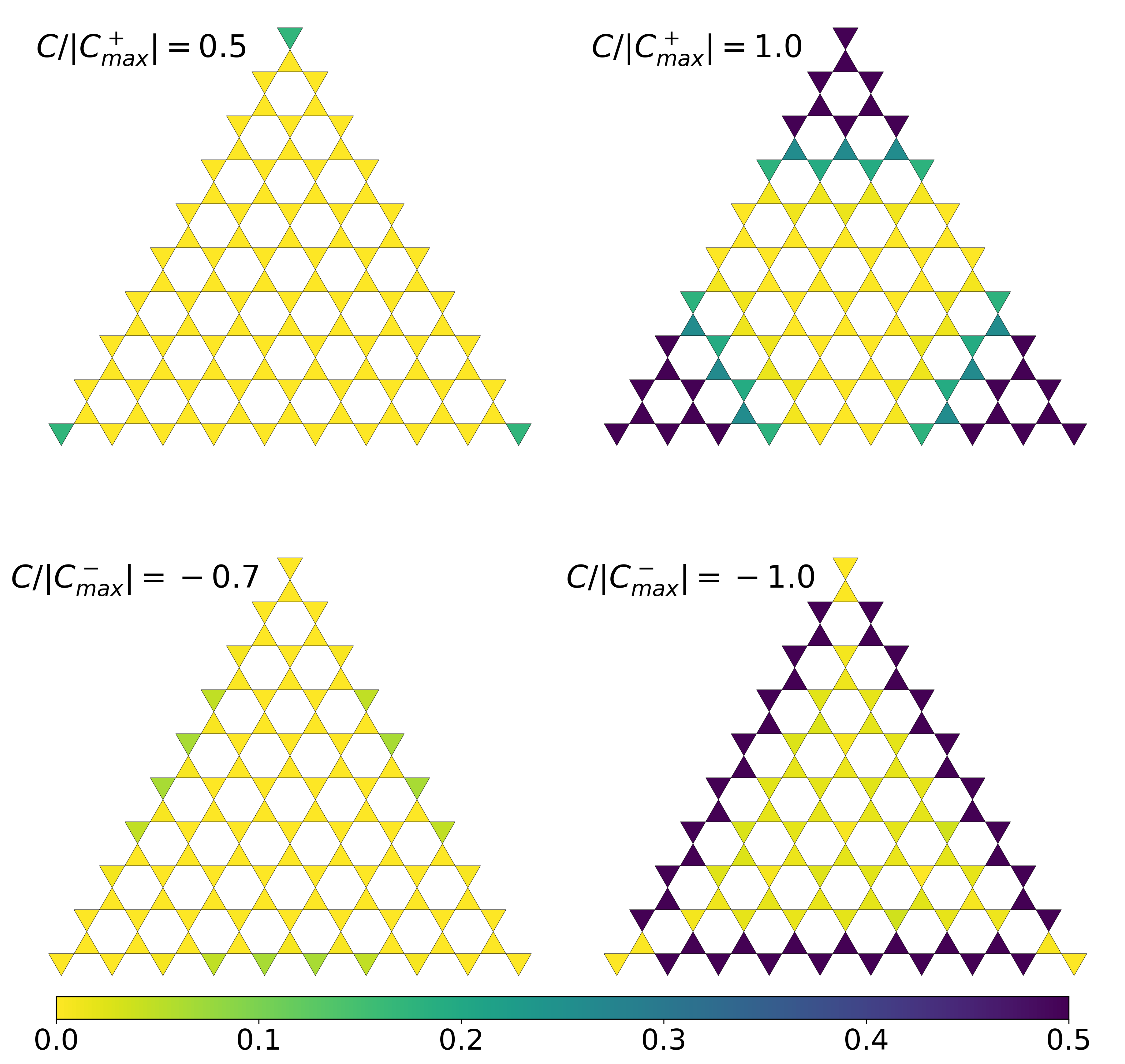}
	\caption{
		Spatial distribution of $|\vec{L}_\alpha|$ in the ground state for systems of size $N=10$ with c-type edges, $\beta=1$, and different values of strain, $C/\Cmax$.
	}
	\label{fig:trisum1}
\end{figure}

The spatial profile of constraint satisfiability can be visualized by plotting the quantity $|\vec{L}_\alpha|$ for each triangle $\alpha$ in the ground state, this is in Fig.~\ref{fig:trisum1}.
With increasing positive strain, the constraint is first violated in the sample corners. For samples with c-type edges, we are able to find an analytic expression in the limit of $\beta\to\infty$ which reads $\Ccr^+ = \sqrt{3} / (10N-9)$. Hence, in this limit we have $\Ccr^+/\Cmax^+ = (4N-3) / (10N-9)$ which tends to $0.4$ for $N\to\infty$. For ic-type edges
we similarly find $\Ccr^+ = \sqrt{3} / (10N-21)$ for $\beta\to\infty$, such that  $\Ccr^+/\Cmax^+ = (4N-9)/(10N-21)$ which again tends to $0.4$ for $N\to\infty$.
For $C<0$, the constraint can be satisfied in the corner triangles for any strain up to $\Cmax$. However, the constraint gets first violated for the triangles in the middle of the boundaries. While we have not been able to obtain a closed-form expression, our numerical evaluation shows that $\Ccr^-/\Cmax^- \to 0.4$ for $\beta\to\infty$ and $N\to\infty$ for both c-type and ic-type edges, as in the case of positive strain, Fig.~\ref{fig:ccr}.

\begin{figure}[!t]
	\includegraphics[width=\linewidth]{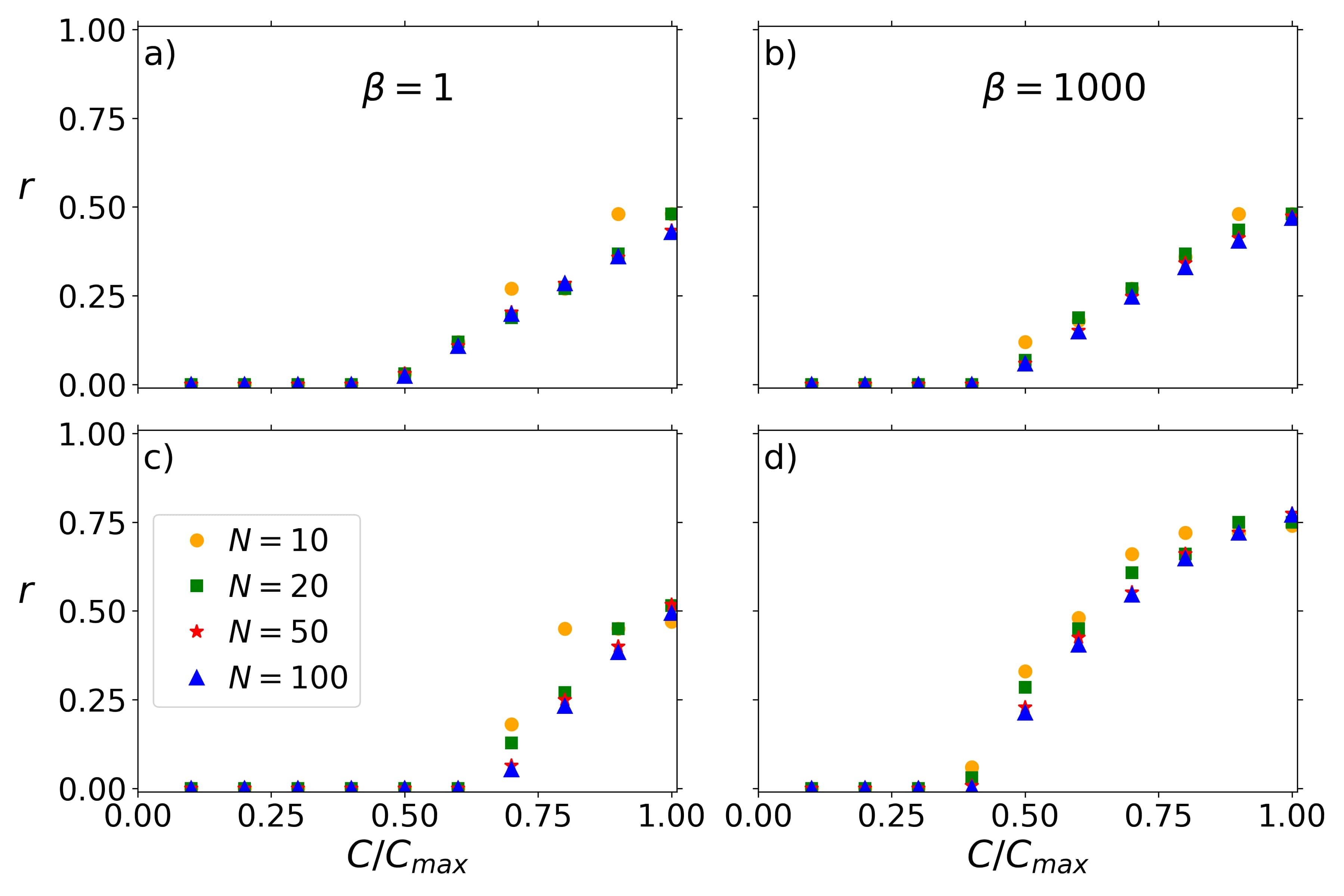}
	\caption{
		Fraction $r\equiv M_{\rm us}/M_{\rm tot}$ of triangles where the constraint $\vec{L}_\alpha=0$ cannot be satisfied in the ground state, plotted as function of $C/\Cmax$ for different $\beta$
		for $C>0$ (a,b) and $C<0$ (c,d). The results are for samples with c-type boundaries; those for ic-type boundaries are similar.
	}
	\label{fig:satisf1}
\end{figure}

For $|C|>|\Ccr|$ the strained system contains multiple triangles where the constraint $\vec{L}_\alpha=0$ cannot be satisfied. If we define the number of these triangles as $M_{\rm us}$, we can consider its ratio with the total number of triangles $M_{\rm tot}$ (which is $N_s/3$ for c-type edges). Apparently $r\equiv M_{\rm us}/M_{\rm tot}$ is zero (non-zero) for $|C|<|\Ccr|$ ($|C|>|\Ccr|$), respectively. In the limit of large system size, $N\to\infty$, $r$ approaches a finite value which remains smaller than unity for $|C|=|\Cmax|$ as triangles near the center of the sample remain weakly distorted even in this limit. Numerical results for $r$ are shown in Fig.~\ref{fig:satisf1}.

Parenthetically, we note that for $|C|<|\Ccr|$ all constraints $\vec{L}_\alpha=0$ can be satisfied, but the ground states cannot be mapped to an origami analog as discussed in Ref.~\onlinecite{origami}, because the geometric condition of Eq.~(2) in that paper is in general not fulfilled by the couplings of the strained system.


\section{Numerical results: Configurations and spin structure factor}
\label{sec:config and SF}

We now turn to our core numerical results, obtained for finite-size triaxially strained kagome-lattice Heisenberg systems using system sizes up to $N=24$.

\subsection{\label{sec:iteration} Iteration scheme}

We use an iterative scheme to find spin configurations corresponding to local minima of the total energy in configuration space. For given values of $N$, $\beta$, and $C$ which determine the Hamiltonian we start from a random initial spin configuration and iteratively minimize the total energy by aligning each spin according to its mean field, supplemented by appropriate random mixing. The iteration is aborted once the average energy per bond, $\varepsilon$, changes less than a threshold $\varepsilon_{\rm conv}$ in one step. For $\varepsilon_{\rm conv}=10^{-8}J$ this happens after typically $10^4 \ldots 10^5$ iteration steps. The iteration is repeated for $N_{\rm init} = 10^5$ different initial conditions. For the state with the globally lowest energy, $E_{\rm min}$, we denote by $\varepsilon_{\rm min}$ its energy per bond, $\varepsilon_{\rm min} = E_{\rm min}/N_b$, where $N_b$ is the number of bonds.
Given the SU(2) spin symmetry of the underlying Hamiltonian, two of the resulting spin configurations are considered equivalent if they match (within a numerical threshold) up to global SU(2) rotations.

Convergence tends to be slow for large systems due to the glassy nature of the energy landscape, see Sec.~\ref{sec:E-landscape} below. Computation time therefore limits our ability to reach larger system sizes, and most calculations are restricted to $N\leq 20$.

\subsection{Ground-state spin configurations}
\label{sec:config}

For any finite strain, $C\neq0$, we find that non-coplanar spin configuration are energetically preferred over coplanar ones: We have verified this tendency by comparing the ground-state energies between those for the SU(2)-symmetric model and models with varying degree of easy-plane anisotropy, obtained by reducing the prefactor of the $S^z S^z$ coupling. The easy-plane models yield a consistently higher ground-state energy, except at zero strain where the ground-state energy does not depend on the anisotropy.

\begin{figure}[t!]
\includegraphics[width=0.95\columnwidth]{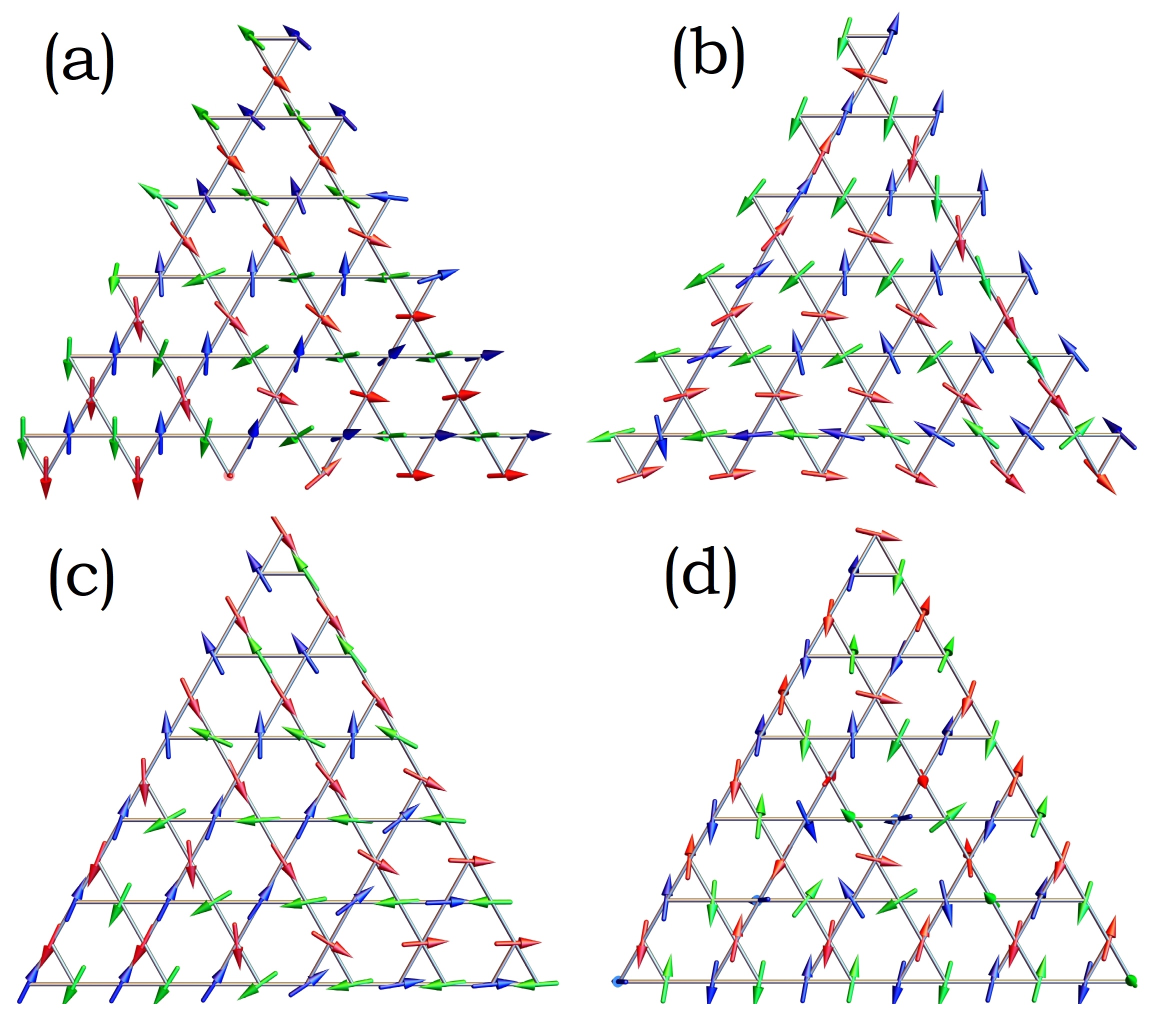}
	\caption{
		Ground-state spin configurations at $|C/\Cmax|=0.95$ for $\beta=1$ and $N=6$ for c-type boundaries: (a) positive and (b) negative strain; and ic-type boundaries: (c) positive and (d) negative strain
	}
	\label{fig:cfg}
\end{figure}

Representative ground-state configurations near maximum strain are shown in Fig.~\ref{fig:cfg}.
While the spin configurations are naturally inhomogeneous, a clear tendency towards local three-sublattice $120^\circ$ order is visible near the sample center; for the unstrained kagome lattice such order is known as $Q=0$ order \cite{harris92}. In contrast, near the sample corners (for $C>0$) or edges (for $C<0$), triangles with ferrimagnetic-like configurations ($\uparrow\uparrow\downarrow$) prevail. This can be rationalized by noting that, in these strongly distorted regions, the spatial distribution of coupling constants, Fig.~\ref{fig:setup}, corresponds to {\em locally} uniaxial strain.

\begin{figure*}[t!]
\includegraphics[width=0.85\textwidth]{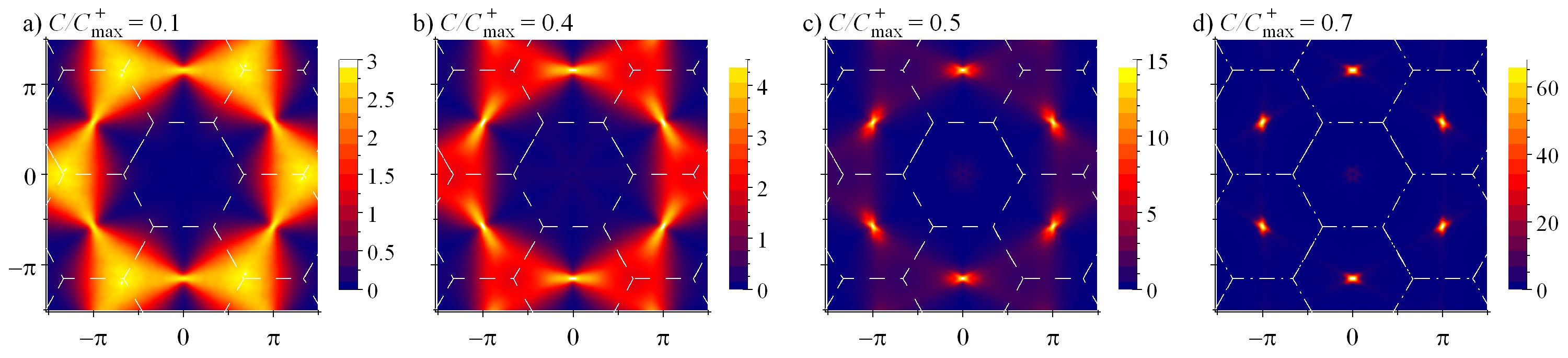}
\includegraphics[width=0.85\textwidth]{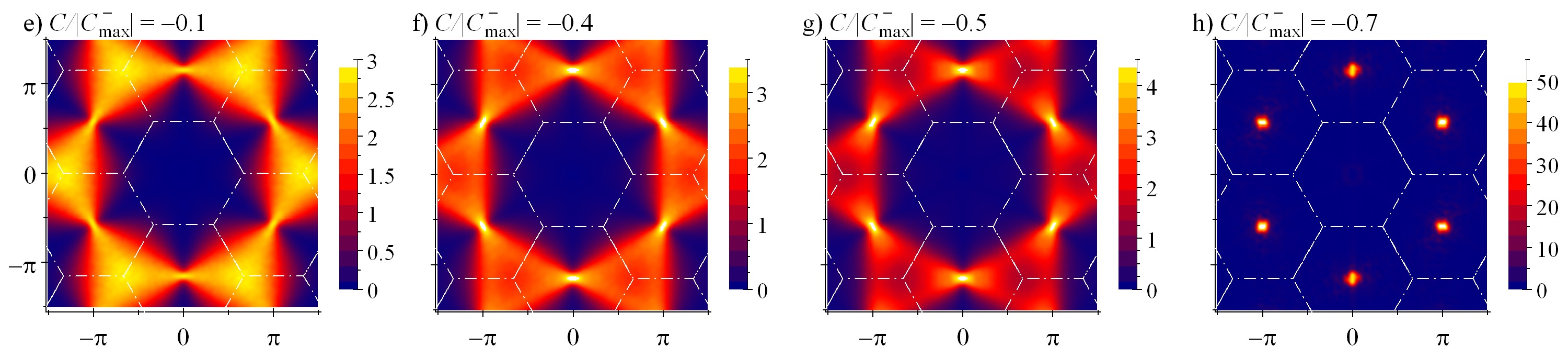}
\caption{
Numerical results for the static spin structure factor $S(\vec q)$, shown as function of $q_x$ and $q_y$ and calculated for $\beta=1$ and samples of size $N=20$ ($N_s=630$) with c-type edges.
(a-d): different values of positive strain $C$.
(e-h): different values of negative strain $C$.
Note that all panels have individual intensity scales. White dashed lines indicate the periodic Brillouin-zone scheme of the kagome lattice.
}
\label{fig:skcol1}
\end{figure*}

In fact, uniaxially strained kagome antiferromagnets have been investigated before \cite{yavorskii07,wang07,schnyder08,nakano11} and display regimes of ferrimagnetism.
Denoting the couplings along one direction by $J$ and along the two others by $J'$, the situation $J\gg J'$ corresponds to chains weakly coupled via middle spins, while the case $J\ll J'$ realizes a square lattice with spin-decorated bonds. In the classical limit, the ground state is a collinear ferrimagnet for $J/J'<1/2$, while for $J/J'>1/2$ there is an infinite family of degenerate canted ferrimagnetic ground states. While the former state is stable also for quantum spins $S=1/2$, the latter is most likely replaced by a spiral state for large $J/J'$ \cite{schnyder08}, and a spin liquid is present in the intermediate regime. The $J/J'<1/2$ ferrimagnet is of obvious relevance to our triaxially strained system, and we will get back to this below.



\begin{figure*}[t!]
\includegraphics[width=0.85\textwidth]{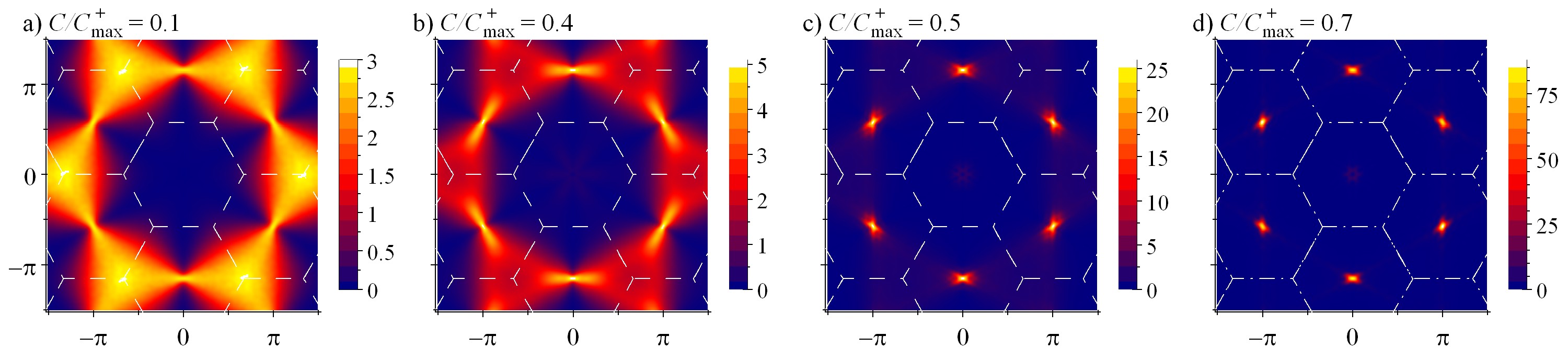}
\includegraphics[width=0.85\textwidth]{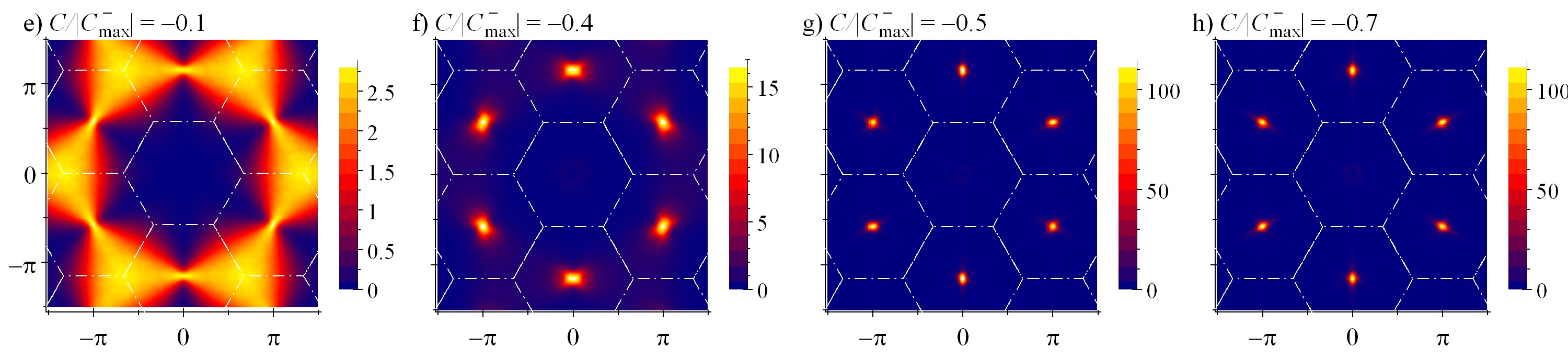}
\caption{
Spin structure factor $S(\vec q)$ as in Fig.~\ref{fig:skcol1}, but now for $\beta=100$. The results are qualitatively similar to that shown in Fig.~\ref{fig:skcol1} for $\beta=1$, with the quantitative differences reflecting the $\beta$ dependence of $\Ccr$.
}
\label{fig:skcol2}
\end{figure*}

\subsection{\label{sec:structure factor} Spin structure factor}

We have analyzed the strain-induced states quantitatively by determining their static spin structure factor, defined as
\begin{equation}
	\label{eq:sq}
	S(\vec q) = \frac{1}{N_s} \sum_{ij} \langle\vec S_i \cdot \vec S_j\rangle e^{i\vec q \cdot (\vec R_i-\vec R_j)}
\end{equation}
where $\langle\ldots\rangle$ denotes an average over the ground-state manifold. For a state with magnetic LRO at wavevector $\vec Q$, the value of $m^2 = S(\vec Q)/N_s$ corresponds to the squared order parameter $m$ in the thermodynamic limit.

In practice, we use $N_{\rm init}$ different initial conditions to find local minima in the energy landscape as described in Sec.~\ref{sec:iteration}. For each converged configuration, we determine the average energy per bond, $\varepsilon=E/N_b$. The averaging in Eq.~\eqref{eq:sq} is then performed over those of the $N_{\rm init}$ converged states whose energy per bond $\varepsilon$ falls in the window $[\varepsilon_{\rm min},\varepsilon_{\rm min}+\Delta\varepsilon]$. The finite energy window $\Delta\varepsilon$ accounts for inaccuracies in convergence and may be interpreted in terms of a finite temperature; it is chosen sufficiently small, $\Delta\varepsilon=10^{-6}J$, unless noted otherwise.

For c-type edges this procedure ensures averaging over a representative set of configurations from the continuously degenerate ground-state manifold. For ic-type edges the sampling is over the ground states plus a small number of low-energy excited states to improve statistics due to the glassy energy landscape which is characterized by many local minima close to the ground-state energy, see Sec.~\ref{sec:E-landscape} below. In calculating $S(\vec q)$ we have varied $\Delta\varepsilon$ for selected parameter sets and found that choosing smaller $\Delta\varepsilon$ changes the values of $S(\vec q)$ by less than 5\% for the system sizes used.

\begin{figure*}[t!]
\includegraphics[width=0.85\textwidth]{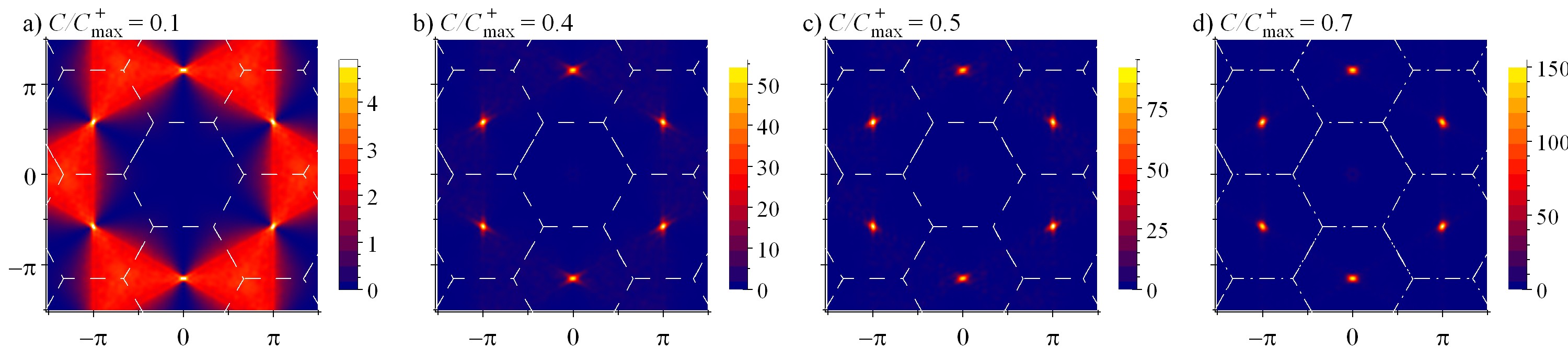}
\includegraphics[width=0.85\textwidth]{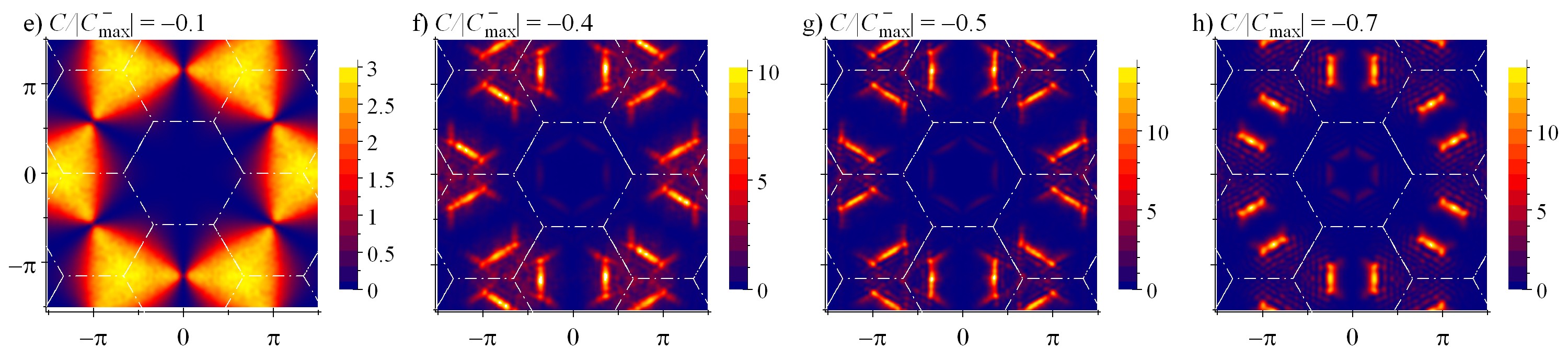}
\caption{
Spin structure factor $S(\vec q)$ as in Fig.~\ref{fig:skcol1}, but now for samples with ic-type edges and $\beta=100$. For positive strain (top) the results are qualitatively similar to that for c-type edges, Fig.~\ref{fig:skcol2}, apart from a slightly smaller value of $\Ccr$. In contrast, for negative strain incommensurate correlations dominate for $|C|>|\Ccr^-|$; those become commensurate only for larger systems, as shown in Fig.~\ref{fig:skcol4} below.
}
\label{fig:skcol3}
\end{figure*}

In general, the calculated spin structure factor displays qualitatively similar behavior for positive and negative strain, for different values of $\beta$, and for different sample edges, as illustrated in Figs.~\ref{fig:skcol1}, \ref{fig:skcol2}, and \ref{fig:skcol3}. At small strain, Figs.~\ref{fig:skcol1}(a,e) and \ref{fig:skcol2}(a,e), the structure factor has the broad shape familiar from the classical kagome Heisenberg model \cite{garanin99,zhito08}, with pinch points located at reciprocal wavevectors $\vec{Q}=\Gamma'$, the centers of higher Brillouin zones, characteristic of the U(1) spin liquid. These pinch points remain sharp under strain, but gain weight with increasing strain, Fig.~\ref{fig:skcol1}(b,f).
At larger strain, Fig.~\ref{fig:skcol1}(c,d,h), pronounced peaks at $\Gamma'$ emerge, which grow with increasing $|C|$ and correspond to three-sublattice $Q=0$ order. Differences between positive and negative strain appear minor. Individual differences, e.g. between Figs.~\ref{fig:skcol2}(b) and (f), can be attributed to the different values of $C/\Ccr$ which can be read off from Fig.~\ref{fig:ccr}.

\begin{figure}[t!]
\includegraphics[width=\columnwidth]{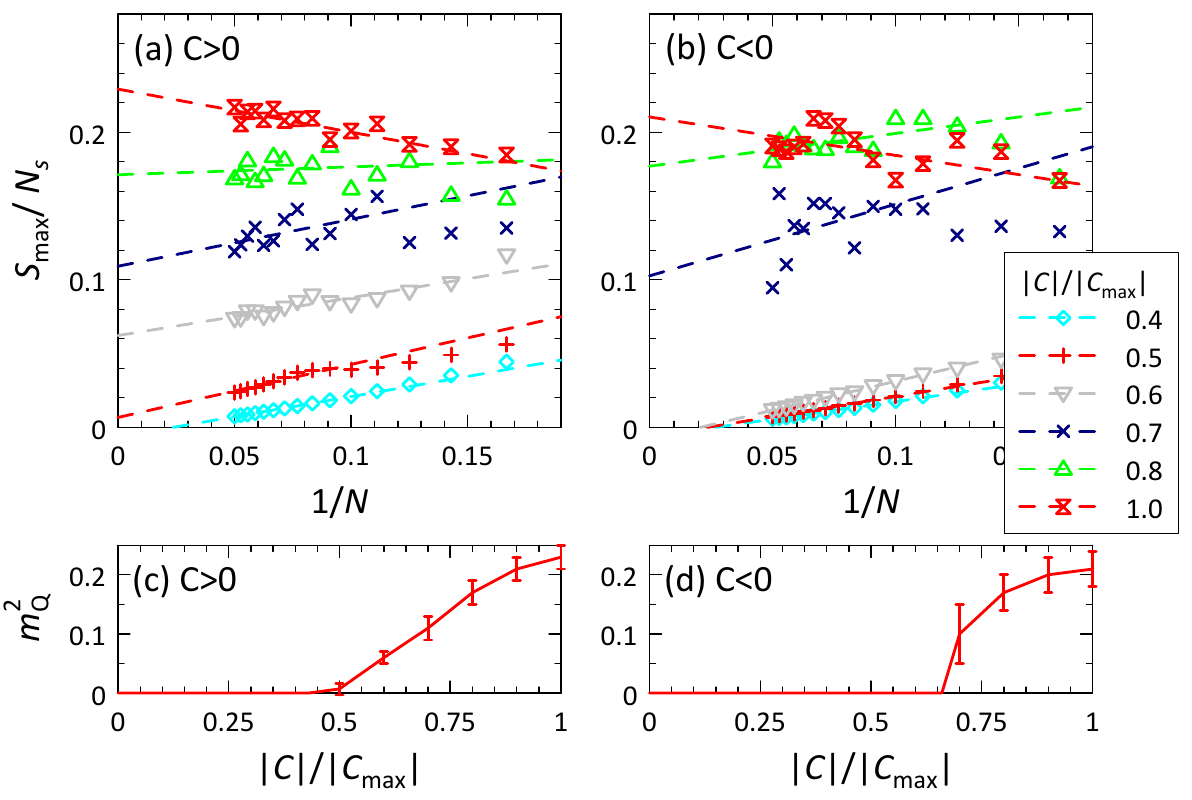}
\caption{
(a,b): Finite-size scaling of $S(\vec Q)/N_s$, corresponding to the Bragg-peak intensity in the static structure factor, for samples with c-type edges and $\beta=1$.
(c,d): Extrapolated peak height as function of $|C|/|\Cmax|$. These results indicate order-parameter-like behavior with a transition located at $\Ccr$.
Panels (a,c) correspond to positive and (b,d) to negative strain.
}
\label{fig:fss}
\end{figure}

\subsection{$Q=0$ order}

Finite-size scaling for the height of the peaks in $S(\vec q)$ at $\vec q=\Gamma'$ and fixed $C/\Cmax$ is demonstrated in Figs.~\ref{fig:fss}(a,b) for $\beta=1$. The data clearly show that $S(\vec Q)/N_s$ scales to zero as $N\to\infty$ for small $|C|$, but tends to a finite value at larger $|C|$. This signals the existence of a magnetically ordered state at large $|C|$, with the transition being located at $C=\Ccr$ within our accuracy, Fig.~\ref{fig:fss}(c,d). Consistent with this, we find that the width of the peaks in $S(\vec q)$ scales to zero as $N\to\infty$ for $|C|>|\Ccr|$.

\begin{figure}[t!]
	\includegraphics[width=\columnwidth]{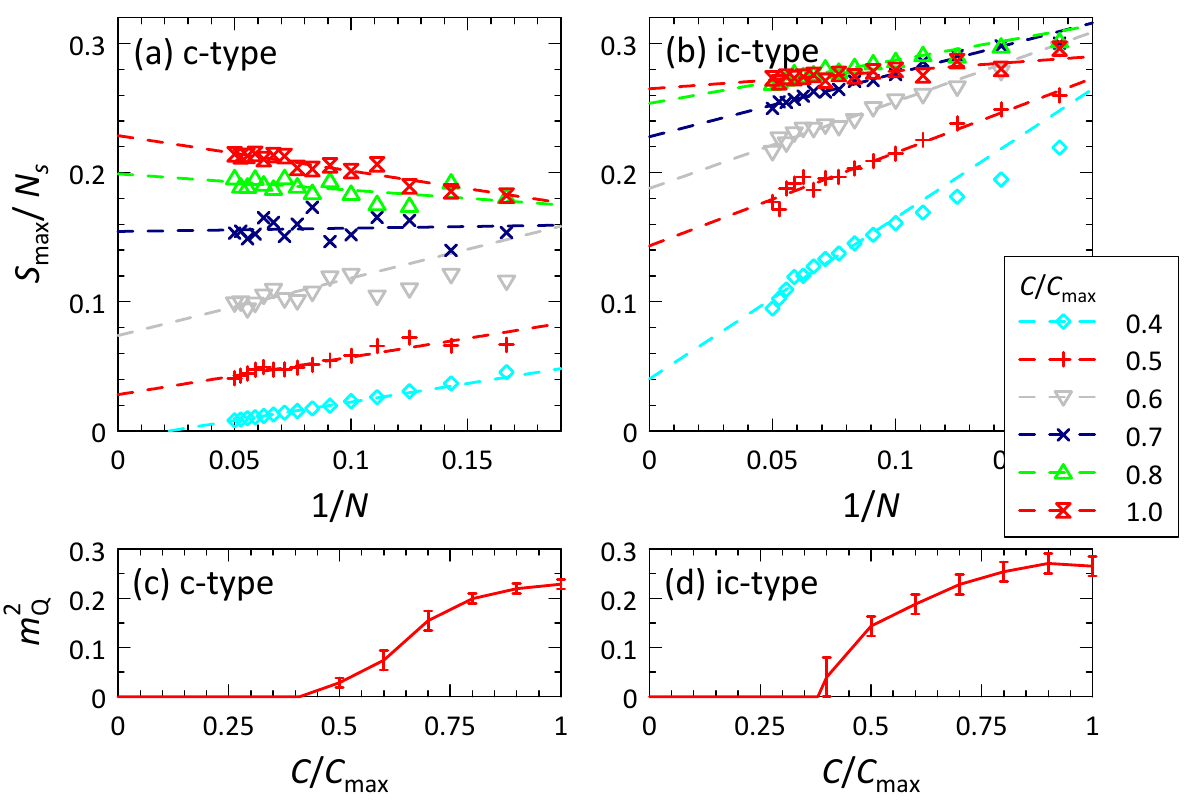}
	\caption{
		(a,b) Finite-size scaling of $S(\vec Q)/N_s$ for $\beta=100$ and (c,d) extrapolated peak height as function of $C/\Cmax$ for samples with $\beta=100$, comparing (a,c) c-type and (b,d) ic-type edges.
	}
	\label{fig:fss2}
\end{figure}

Fig.~\ref{fig:fss} also illustrates non-monotonic $N$ dependencies which can be traced back to commensurability effects of inhomogeneous spin arrangements near the sample boundaries. A finite-size-scaling comparison between samples with c-type and ic-type edges for positive strain is shown in  Fig.~\ref{fig:fss2}, here for $\beta=100$. While finite-size systems with ic-type edges tend to have a larger order parameter and a smaller non-monotonic $N$ dependence, the extrapolated data show only minor differences, partially related to the different $C/\Ccr$. Together, this underlines that the $Q=0$ order is a robust bulk phenomenon, with boundary effects being subleading.

To rationalize the appearance of the $Q=0$ order, we recall that the triaxially strained kagome flake realizes bond configurations at the corners (at the mid-edges) for $C>0$ ($C<0$) which correspondi to a uniaxially strained systen with $J/J'<1/2$. The latter displays ferrimagnetic order \cite{yavorskii07,wang07,schnyder08,nakano11} with magnetic Bragg peaks at two of the $\Gamma'$ momenta, namely those perpendicular to the $J$ bond direction. Therefore, the six $\Gamma'$ peaks present for $|C|>|\Ccr|$ in the triaxially strained sample can be thought of as a superposition of three domains of ferrimagnetic configurations.
We note, however, that this picture is oversimplified, as (i) the entire sample contributes to the Bragg peaks and (ii) the local spin configurations deviate significantly from the collinear ferrimagnet.

\begin{figure*}[t!]
\includegraphics[width=0.85\textwidth]{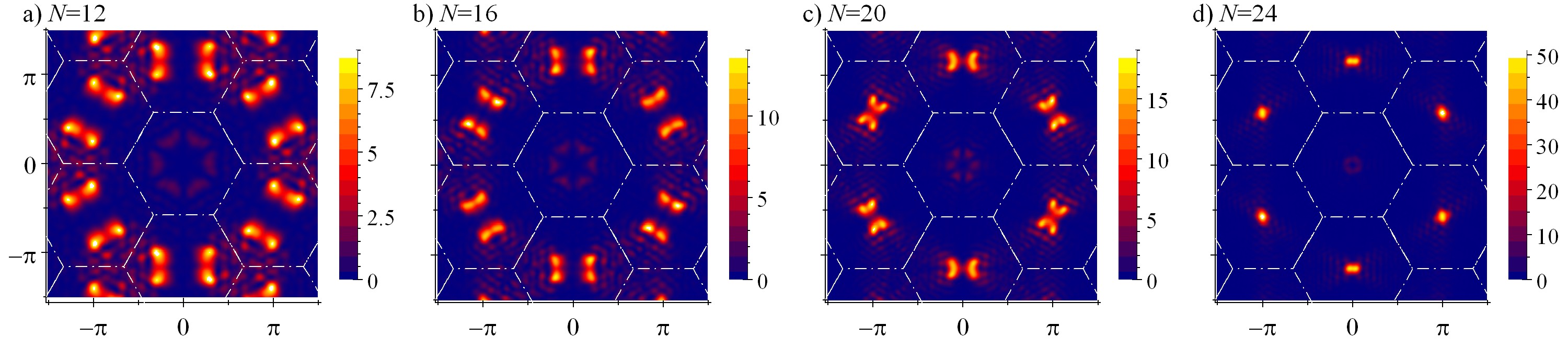}
\caption{
Spin structure factor $S(\vec q)$ for samples with ic-type edges, $C/|\Cmax^-|=-0.8$, and $\beta=100$. Results are shown for linear system sizes (a) $N=12$, (b) $16$, (c) $20$, and (d) $24$, demonstrating the evolution from peaks at incommensurate locations to commensurate ones with increasing system size, for details see text.
}
\label{fig:skcol4}
\end{figure*}

A particular situation arises for samples with ic-type edges and $C<0$: Here, the sample edges tend to have configurations of the form $\uparrow\uparrow\downarrow\downarrow$, Fig.~\ref{fig:cfg}(d), because -- compared to c-type edges -- the outer triangles with strong bonds have been removed, producing pairs of effectively strongly coupled spins along the edge. The $\uparrow\uparrow\downarrow\downarrow$ configurations in turn lead to structure-factor peaks at incommensurate locations, see Fig.~\ref{fig:skcol3}(f-h). As these configurations are restricted to the boundary row, their influence on the structure factor diminishes with increasing $N$, such that for sufficiently large samples commensurate peaks in $S(\vec q)$ are restored, Fig.~\ref{fig:skcol4}.

\begin{figure}[t!]
	\includegraphics[width=0.87\columnwidth]{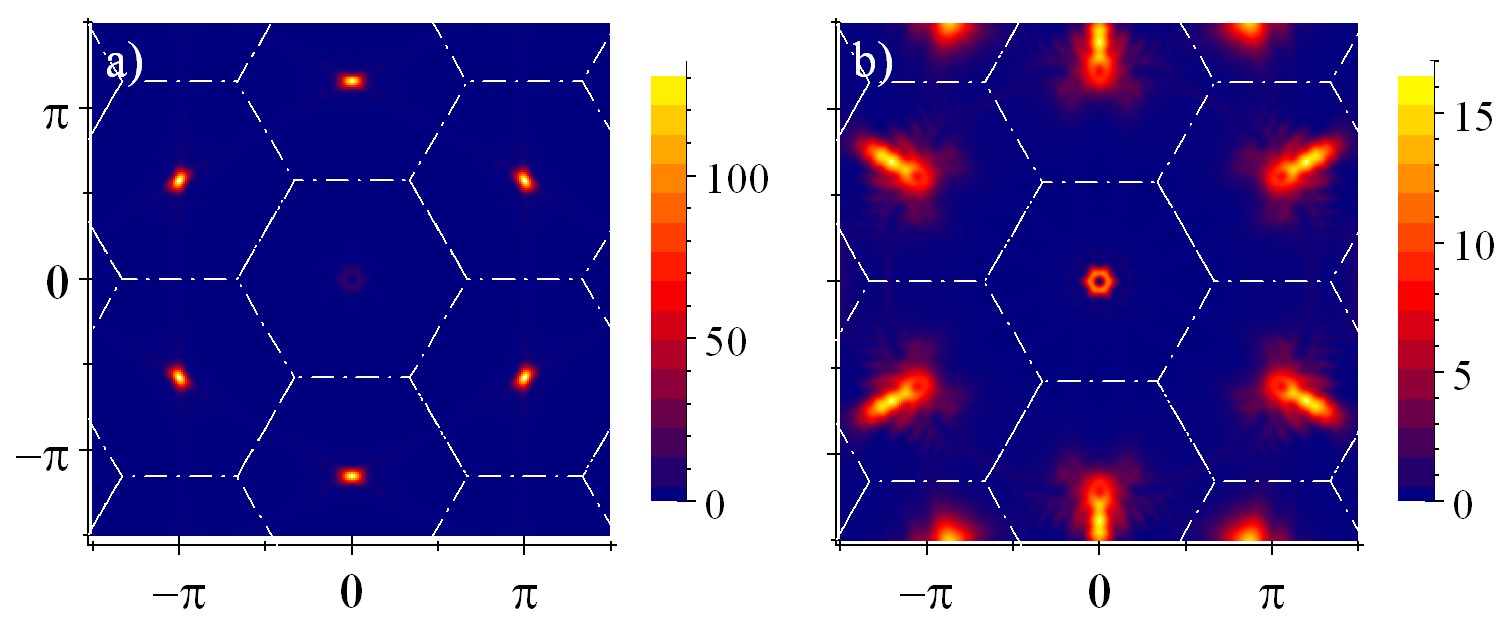}
	\caption{
		Comparison of the spin structure factor calculated with (a) unstrained and (b) strained coordinates for $\beta=3$, shown for $N=20$ samples with c-type edges and maximum positive strain, $C=\Cmax^+$.
	}
	\label{fig:skcol5}
\end{figure}

For strongly distorted lattices, it makes a difference whether the lattice coordinates used to calculate $S(\vec q)$ in Eq.~\eqref{eq:sq} are taken as the unstrained $\vec{R}_i$ or the strained $\vec{R}_i+\vec{U}_i$; a neutron-scattering experiment would probe the latter. For simplicity, Figs.~\ref{fig:skcol1}-\ref{fig:skcol4} have been calculated with unstrained coordinates. A comparison of the structure factors calculated with both unstrained and strained coordinates is shown in Fig.~\ref{fig:skcol5} for a large-strain case and a realistic value of $\beta=3$: One sees that the strain-induced change in lattice geometry leads to a broadening of the Bragg peaks. For larger $\beta$ the geometric lattice distortion at a given $C/\Cmax$ is smaller and hence the difference between the two cases diminishes.

\begin{figure}[t!]
	\includegraphics[width=0.45\columnwidth]{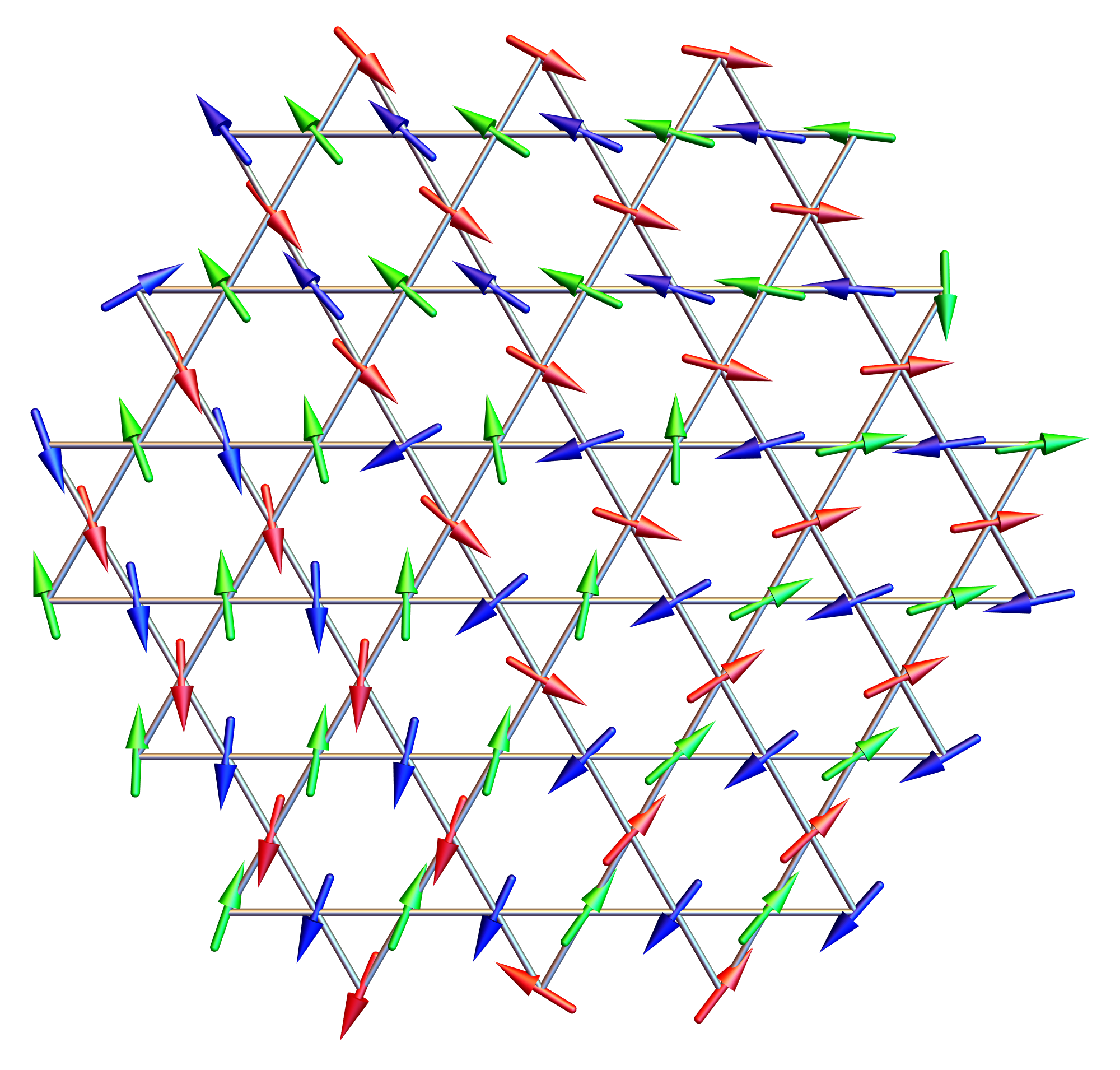}\hfill%
	\includegraphics[width=0.47\columnwidth]{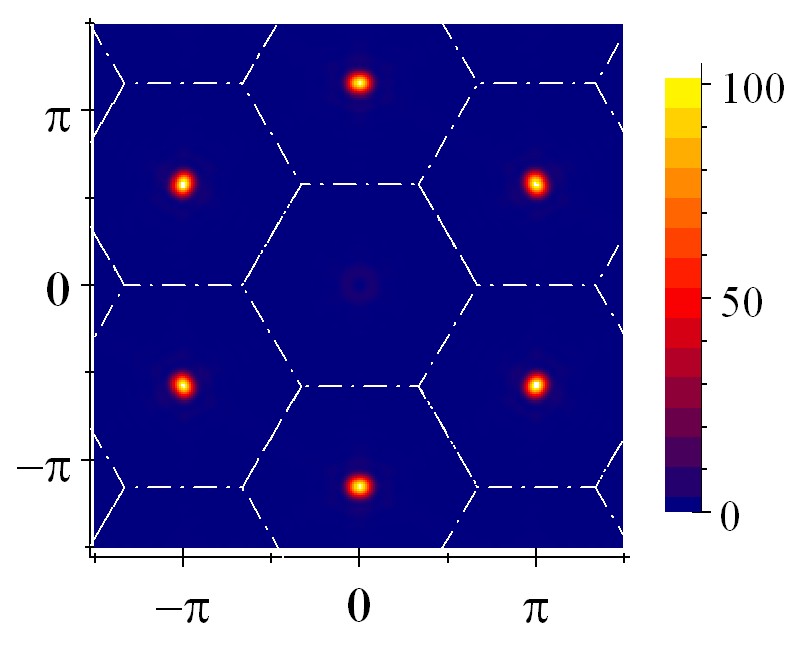}
	\caption{
		Results for hexagonal-shaped kagome flakes with c-type edges subject to triaxial strain.
		Left: Spin configuration for a system with $N_s=90$ sites and $C=0.95 \Cmax$.
		Right: Spin structure factor for $C/\Cmax=0.7$, $\beta=1$, and $N_s=462$ sites.
	}
	\label{fig:hex}
\end{figure}

For completeness we have also considered kagome flakes of shapes different from triangular, in particular hexagonal and circular. For sufficiently large samples, the structure-factor results are similar to those for triangular-shaped samples. We illustrate this in Fig.~\ref{fig:hex} which shows spin configuration and structure factor for a hexagonal-shaped system under triaxial strain. We note that, for such samples, positive and negative strain are equivalent by symmetry.


\section{Glassy energy landscape}
\label{sec:E-landscape}

Our iterative minimization scheme keeps track of a large set of local minima in the energy landscape. For the problem at hand, we hand found the latter to be surprisingly complex.

\subsection{\label{sec:local minima} Local minima and their energy distribution}

For very small systems, $N\lesssim 5$, we find that essentially all iterations converge to states with the same lowest energy. Inspecting the spin configurations for $|C|>|\Ccr|$ indicates the existence of few discrete degenerate ground states for ic-type boundaries, whereas systems with c-type boundaries appear to display continuously degenerate ground states (except at $C=\Cmax$), as repeated iterations find inequivalent states with identical energies.

\begin{figure}[tb]
\includegraphics[width=\linewidth]{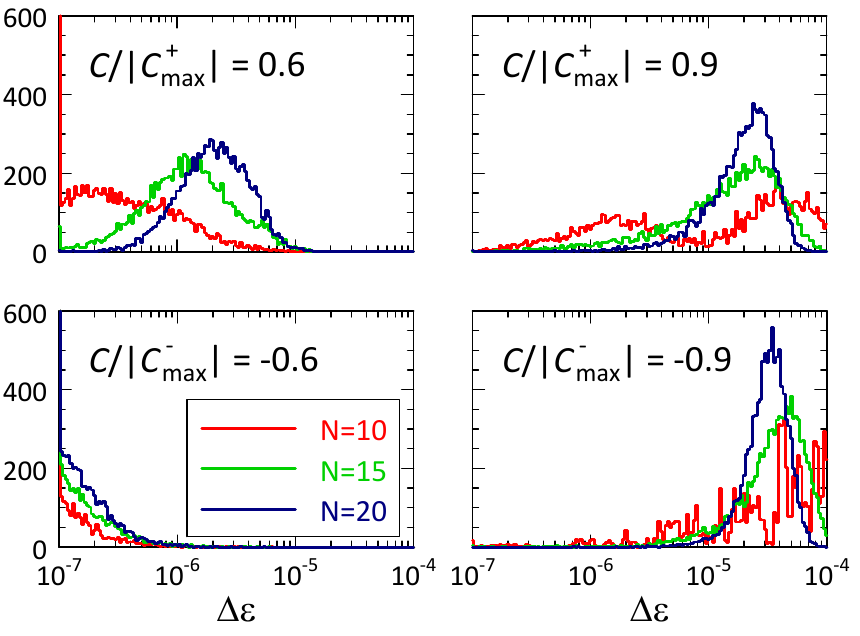}%
\caption{
Histograms of converged energies per bond, $\varepsilon$, representing local minima of the energy landscape. Data are shown for different values of $C/\Cmax$ and different system sizes $N$ for samples with c-type boundaries and $\beta=1$. The horizontal axis shows the energy relative to the global minimum, $\Delta\varepsilon = \varepsilon - \varepsilon_{\rm min}$, on a logarithmic axis, the vertical axis the frequency out of $10^4$ random initial conditions.
}
\label{fig:ehist1}
\end{figure}

The situation is drastically different for larger systems. While convergence to the same lowest energy is still common for $|C|<|\Ccr|$, the iteration scheme finds local minima with widely distributed energies for $|C|>|\Ccr|$. Sample distributions for the energy per bond, $\varepsilon-\varepsilon_{\rm min}$, are shown in Fig.~\ref{fig:ehist1}: The distributions appear effectively continuous for large systems, with a width reaching up to $10^{-4}J$. Naturally, the distribution width increases with increasing strain $C$, while for $C\to 0$ the width tends to zero.

These results signal a complex energy landscape for $|C|>|\Ccr|$, with abundant local minima which are separated by energy barriers. This behavior is well known for spin glasses where randomness and frustration conspire to produce complex low-energy states without long-range order \cite{fischer}. Remarkably, we find a glassy energy landscape here in a system free of quenched disorder (but with inhomogeneous couplings), and our structure-factor results show that this glassy behavior coexists with signatures of magnetic long-range order.


\subsection{Ground-state degeneracy}
\label{sec:GS degeneracy}

We made an attempt to estimate the ground-state degeneracy -- up to global SU(2) rotations -- by monitoring the number of \emph{different} \cite{diffnote} converged spin configurations, $N_{\rm diff}$, whose energy equals the minimum energy within a small window $\Delta\varepsilon=10^{-9}J$.
For samples with c-type boundaries we find that $N_{\rm diff}$ scales with $N_{\rm init}$ for all values of $C$ and system sizes $N$, indicating that the ground states are continuously degenerate. This is consistent with the finding of non-trivial zero modes reported below.

\begin{table}[tb]
\centering
\begin{tabular}{m{1cm}||m{0.7cm}|m{0.7cm}|m{0.7cm}|m{0.7cm}||m{0.7cm}|m{0.7cm}|m{0.7cm}|m{0.7cm}}
	Size & \multicolumn{8}{c}{$C/|\Cmax|$} \\
	$N$  & $0.5$ & $0.6$ & $0.7$ & $0.8$ & $-0.5$ & $-0.6$ & $-0.7$ & $-0.8$ \\ \hline\hline
	6    & 3   & 3   & 1   & 1   & 380  & 396  & 1025 & 7    \\ \hline
	8    & 1   & 2   & 1   & 4   & 26   & 20   & 306  & 51   \\ \hline
	10   & 1   & 1   & 1   & 1   & 17   & 2    & 52   & 15   \\ \hline
\end{tabular}
\caption{
	Number of different degenerate lowest-energy states for samples with ic-type edges and $\beta=1$ for different $N$ and $C/\Cmax$, obtained from $N_{\rm init}=10^5$ initial configurations, for details see text.}
\label{table:gs}
\end{table}
In contrast, for samples with ic-type boundaries we observe that $N_{\rm diff}$ tends to saturate with increasing $N_{\rm init}$, at least for strain values $|\Ccr| < |C| < |\Cmax|$. The actual number of ground states depends on both $C$ and $N$, with non-monotonic variations, see Table~\ref{table:gs}.
We point out, however, that counting \emph{true} ground states is a numerically expensive task for large systems due to the glassy nature of the energy landscape, and a compromise between runtime, convergence accuracy, and selection window $\Delta\varepsilon$ is required. Based on the available data, we are not able to determine how the ground-state degeneracy of ic-type samples scales with system size; the results clearly point toward a non-extensive number.


\subsection{\label{sec:hessian} Low-energy modes}

To further characterize the states of the strained kagome Heisenberg magnet, we determine the quadratic energy cost of fluctuations around ground-state configurations. To this end, we construct the Hessian matrix for a given minimum-energy state and determine its eigenvalues and eigenvectors.

The Hessian is constructed as described in Ref.~\onlinecite{bilitewski19}: For a spin configuration $\lbrace \textbf{s}_i \rbrace$ we choose an orthonormal local basis $\left(\textbf{s}_i,\textbf{u}_i ,\textbf{v}_i \right)$ at every lattice site and parameterize fluctuations as $ \tilde{\textbf{s}}_i = \sqrt{1- \epsilon_i^2} \textbf{s}_i  + \epsilon_{ui}\textbf{u}_i + \epsilon_{vi}\textbf{v}_i $ with $\epsilon_i = (\epsilon_{ui}, \epsilon_{vi}) $ which takes into account the spin normalization condition. The quadratic energy cost of fluctuations around a ground state is given by $E = \epsilon^T M \epsilon $ where $M$ is the Hessian matrix with dimension $2N_s \times 2N_s$. Diagonalizing the Hessian matrix gives us the eigenvalues $\lambda_j$ and the corresponding eigenvectors.

\begin{table}[!b]
	\centering
	\begin{tabular}{m{1cm}||m{0.7cm}|m{0.7cm}|m{0.7cm}|m{0.7cm}||m{0.7cm}|m{0.7cm}|m{0.7cm}|m{0.7cm}}
		Size & \multicolumn{8}{c}{$C/|\Cmax|$} \\
		$N$  & $0.1$ & $0.4$ & $0.7$ & $1.0$ & $-0.1$ & $-0.4$ & $-0.7$ & $-1.0$ \\ \hline\hline
		6    & 18  & 18  & 6   & 3   & 18   & 18   & 6    & 9    \\ \hline
		8    & 24  & 24  & 12  & 3   & 24   & 24   & 9    & 9    \\ \hline
		10   & 30  & 30  & 10  & 3   & 30   & 30   & 10   & 9    \\ \hline
	\end{tabular}
	\caption{
		Number of Hessian zero modes for samples with c-type edges and $\beta=1$ for different $N$ and $C/\Cmax$, for details see text.
	}
	\label{table:zero modes}
\end{table}

The spectrum of the Hessian always contains three trivial zero modes (Goldstone modes) due to the SU(2) symmetry of the underlying Hamiltonian.
Interpreting all eigenvalues below $10^{-6}J$ as zero modes, we find that samples with c-type boundaries generically display additional, i.e., non-trivial zero modes implying a continuous ground-state degeneracy, except for $C=\Cmax^+$. The zero-mode count is shown Table~\ref{table:zero modes}. The number of zero modes increases with linear system size $N$ for $|C|<|\Ccr|$, while it appears to saturate for $|C|>|\Ccr|$. We have analyzed the zero-mode eigenvectors by calculating their inverse participation ratio and by inspecting their spatial distribution (not shown) and concluded that the non-trivial modes primarily live near the c-type sample boundaries for any non-zero strain.
Boundary-induced zero modes are in fact consistent with the analysis of bond-disordered kagome antiferromagnets in Ref.~\onlinecite{bilitewski17} which concluded that no zero modes should exist in the bulk for inhomogeneous distributions of magnetic couplings.

In contrast, samples with ic-type boundaries do not feature non-trivial zero modes, except at $C=\Cmax^-$ where the corner spins are disconnected and can be trivially rotated. This implies that the ground states display discrete degeneracies only (apart from global rotations), again consistent with Ref.~\onlinecite{bilitewski17}.

\begin{figure}[tb!]
	\includegraphics[width=\linewidth]{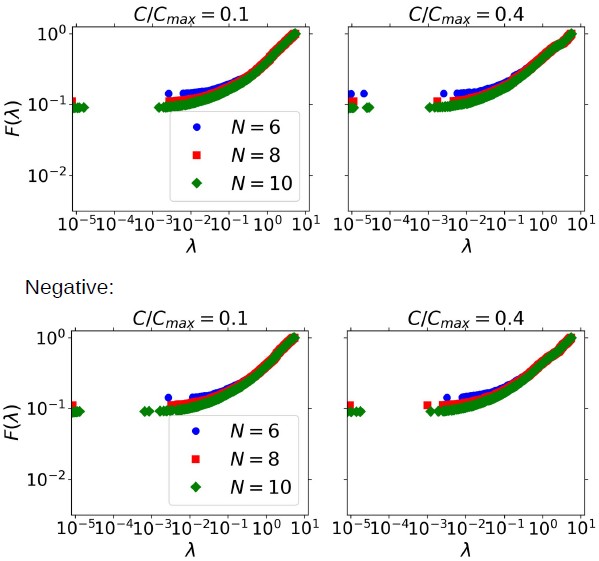}
	\caption{
		Cumulative distribution function $F(\lambda$) of the Hessian eigenvalues, plotted on a log-log scale, for different values of (a,b) positive and (c,d) negative strain. The data have been obtained for samples with different $N$, $\beta=1$, c-type edges, and averaged over 10 different ground-state configurations.
	}
	\label{fig:hess_eval_c}
\end{figure}

\begin{figure}[tb!]
	\includegraphics[width=\linewidth]{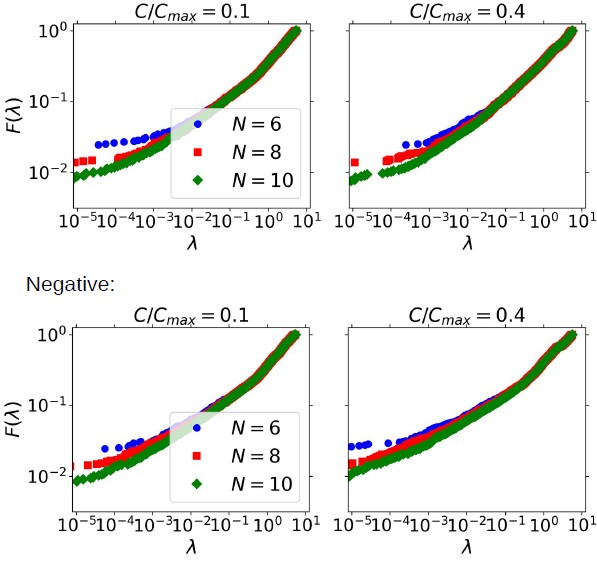}
	\caption{
		Same as Fig.~\ref{fig:hess_eval_c}, but now for samples with ic-type edges.
	}
	\label{fig:hess_eval_ic}
\end{figure}

The character of the finite-energy spectrum can be analyzed via the cumulative distribution function $F(\lambda) = 1/(2N_s) \sum_j \Theta(\lambda-\lambda_j)$ of the Hessian eigenvalues $\lambda_j$ for a particular local minimum. Plots of the cumulative distribution function are shown in Figs.~\ref{fig:hess_eval_c} and \ref{fig:hess_eval_ic}.

Independent of the edges, the spectrum appears gapless for large $N$: existing gaps in the spectrum get filled with increasing $N$, indicating that these are finite-size effects. Interestingly, for c-type (ic-type) edges the density of low-energy modes increases (decreases) with increasing $|C|$. This can be rationalized by considering that, with increasing $|C|$, for $c$-type edges zero modes are converted into low-$E$ modes, whereas for ic-type edges all modes are shifted to higher energy.


\section{Summary and outlook}
\label{sec:conclusion}

Non-uniform strain can be used to drive highly frustrated magnets into novel states: We have demonstrated this for triaxial strain applied to the classical kagome Heisenberg antiferromagnet: While this model system, in the absence of strain, is in a highly degenerate classical spin-liquid state, weak strain partially lifts the degeneracies. The system enters a non-coplanar spin-liquid state with pronounced strain-driven short-range spin correlations. Larger strain drives a phase transition into a state with $Q=0$ long-range order, and we have connected this to the tendency towards ferrimagnetism in the uniaxially strained kagome antiferromagnet. Most interestingly, the inhomogeneously ordered state at large strain displays a rugged energy landscape akin to that of a spin glass.

Our results demonstrate an intriguing coexistence of magnetic long-range order and a glassy energy landscape in a classical non-random spin system. This calls for a deeper understanding of its dynamic properties, not only at the linear-response level, but also concerning relaxation and quenches. In this context, the role of the zero modes present for c-type boundaries needs particular attention.
At finite temperatures, strain effects will compete with thermal order by disorder, which may drive novel types of phase transitions.

A notoriously difficult question is that for quantum effects at $T=0$. Perhaps most interesting is the physics of the strained $S=1/2$ kagome Heisenberg antiferromagnet. Provided that the unstrained system is a topological $Z_2$ spin liquid, it features a gap to $Z_2$ vortex excitations (visons) and hence can be expected to be stable at least against small strain. However, numerics indicates that the unstrained system is sensitive to small perturbations, related to its proximity to one or more transitions between different ground-state phases \cite{sheng15,trebst16,normand17,wietek19}. Hence, moderate strain is likely sufficient to modify the quantum ground state: Given the rugged energy landscape, we believe the strained kagome quantum antiferromagnet presents a fascinating platform to study quantum glassiness and aspects of many-body localization.

Our work suggests to consider strain engineering of degenerate states on a more general level and, in particular, prompts generalizations to other strain patterns as well as highly frustrated magnets on other lattices, such as pyrochlore or hyperkagome. While this is left for future work, we speculate here that large strain generically induces ordering tendencies, and it will be extremely interesting to study the emergence of corresponding ordering transitions, both at zero and finite temperatures.

\begin{figure*}[t!]
\includegraphics[width=0.85\textwidth]{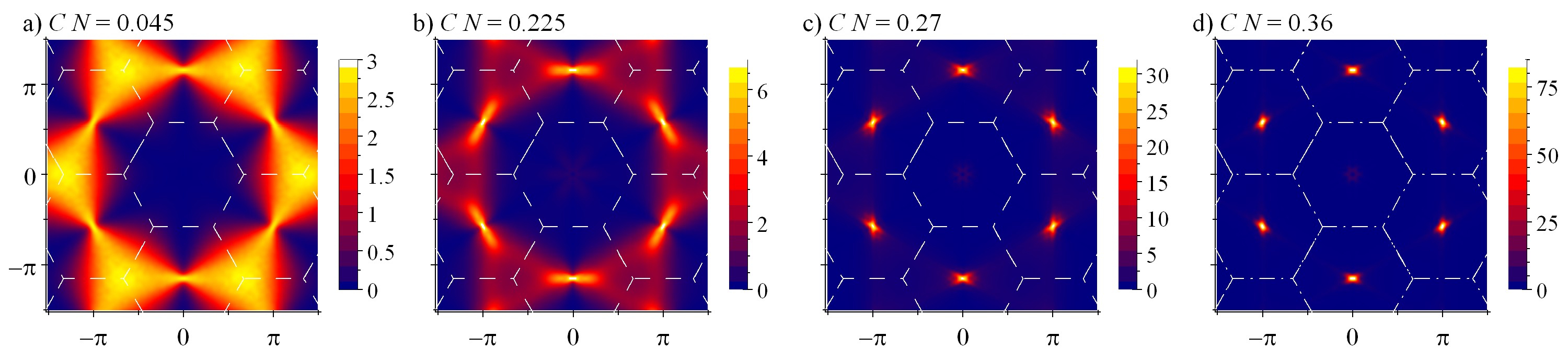}
\caption{
Spin structure factor $S(\vec q)$ as in Fig.~\ref{fig:skcol1}, but here for strained samples with c-type edges, $\beta=3$, and exponential length dependence \eqref{expjij} of coupling constants $J$.  The results are similar to that obtained using the linearized length dependence, Eq.~\eqref{ourjij}.
}
\label{fig:skcol6}
\end{figure*}

\acknowledgments

We thank D. Arovas, L. Fritz, I. G\"othel, R. Moessner, S. Rachel, and S. Trebst for discussions as well as collaborations on related work.
We acknowledge financial support from the Deutsche Forschungsgemeinschaft (DFG) through SFB 1143 (project-id 247310070) and the W\"urzburg-Dresden Cluster of Excellence on Complexity and Topology in Quantum Matter -- \textit{ct.qmat} (EXC 2147, project-id 390858490) as well as by the IMPRS on Chemistry and Physics of Quantum Materials.


\appendix

\section{Exponential vs. linearized bond-length dependence of coupling constants}
\label{sec:expoj}

The results presented in the body of the paper employ the simplified geometry dependence of the magnetic exchange couplings given in Eq.~\eqref{ourjij}. In a real material this dependence is not linear and will in general also depend on bond angles. The angle dependence arises from anisotropic orbitals as well as from superexchange paths via intermediate ions and is clearly material-dependent.

\begin{figure}[tb]
\includegraphics[width=\columnwidth]{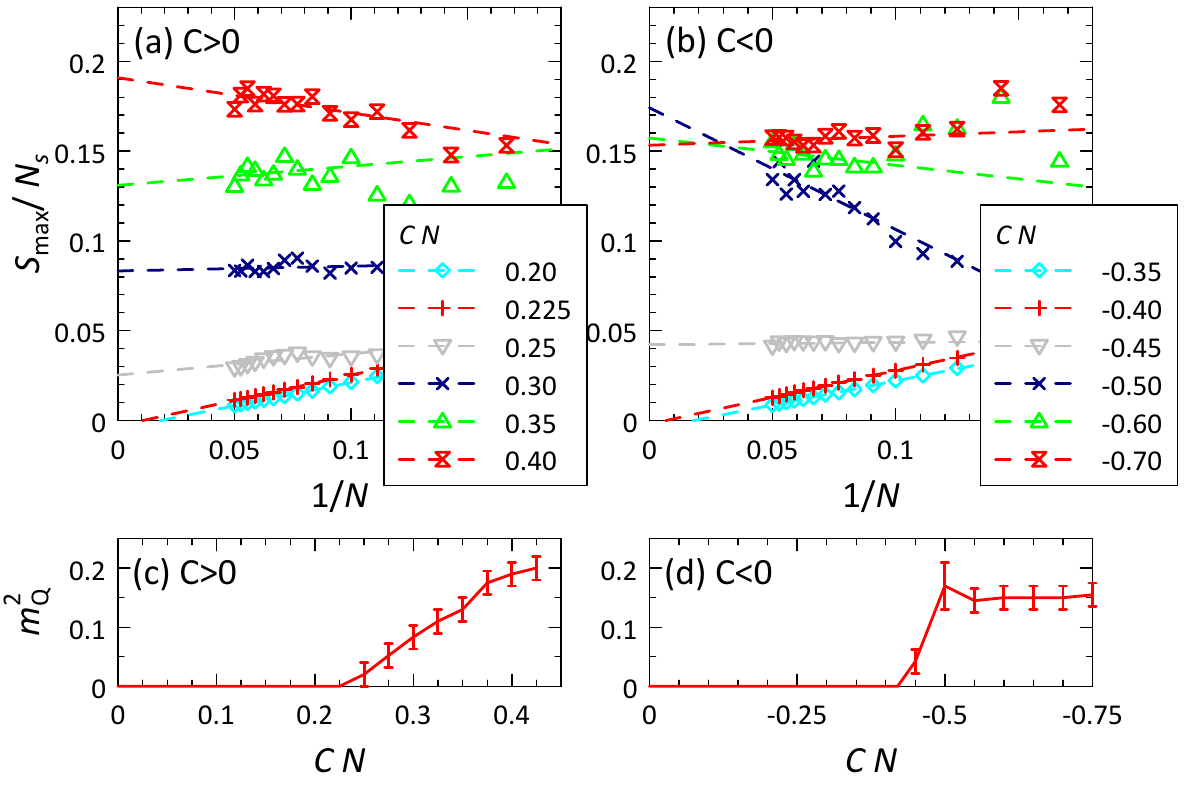}
\caption{
(a,b) Finite-size scaling of $S(\vec Q)/N_s$ and (c,d) extrapolated peak height as function of $C N$ for (a,c) positive and (b,d) negative strain, here for strained samples with c-type edges, $\beta=3$, and {\em exponential} (instead of linearized) length dependence of coupling constants.
}
\label{fig:fss3}
\end{figure}

Here we illustrate the robustness of our findings by assuming an exponential bond-length dependence instead of the linearized one, i.e., we use
\begin{equation}
\label{expjij}
J_{ij} = J \exp\left[- \beta (|\vec{\delta}_{ij}|/a_0 - 1)\right]\,.
\end{equation}
As these couplings are always positive, our previous definition of $\Cmax$ ceases to be well-defined. Therefore, we now use $C N= \bar{C}N\beta a_0$ as size-independent dimensionless measure of the strain effect on the coupling constants.

We have performed numerical simulations using Eq.~\eqref{expjij} instead of Eq.~\eqref{ourjij} and find the results to be qualitatively unchanged: Upon increasing the strain, both the non-coplanar spin liquid and the glassy $\vec Q=0$ ordered state appear in an essentially unchanged fashion. This is illustrated in Figs.~\ref{fig:skcol6} and \ref{fig:fss3}, showing the spin structure factor and the finite-size scaling for its peak height; these figures can be compared to Figs.~\ref{fig:skcol1}, \ref{fig:skcol2}, \ref{fig:fss}, and \ref{fig:fss2}(a,c).

The robustness can be rationalized as follows: Compared to its linearized version, the exponential coupling dependence \eqref{expjij} leads to somewhat larger couplings for long (i.e. weak bonds) and to significantly larger couplings for short (i.e. strong bonds), while undistorted bonds remain unchanged. As a result, elementary triangles whose linearized couplings \eqref{ourjij} strongly violate the triangle inequality $\gamma_{i\alpha}+\gamma_{j\alpha}>\gamma_{k\alpha}$, see Sec.~\ref{sec:critical}, and thus cannot satisfy the constraint $\vec{L}_\alpha=0$, continue to do so for an exponential dependence \eqref{expjij}. As unsatisfied triangles force the  emergence of the ordered glassy phase, its appearance and character remains unchanged. Switching from linearized to exponential coupling dependence shifts the transition location $\Ccr$ by about 15\% (recall, e.g., $\Cmax^+ N = \sqrt{3}/4 \approx 0.433$ for $N\to\infty$).

We conclude that the linearization of the couplings' length dependence, Eq.~\eqref{ourjij}, is a permissable approximation in the regime of interest. We note that the same linearization approximation is frequently used in the literature on strained graphene where it has been shown to
remain reasonably accurate for sample deformations up to 10-15\% \cite{peeters13,settnes16}.



\begin{thebibliography}{}
\bibitem{diffnote}
Comparison of spin configurations is made by comparing all scalar products $\vec{S}_i\cdot\vec{S}_j$; two configurations are considered different if any of the $\vec{S}_i\cdot\vec{S}_j$ differ by more than $\Delta s=10^{-3}$.
\end{thebibliography}
\end{document}